\documentclass[twocolumn,showpacs,amsmatsh,amssymb,aps,prc,groupedaddress,superscriptaddress]{revtex4-1}

\usepackage[caption=false]{subfig}
\usepackage{tabularx}
\usepackage{graphicx}
\graphicspath{{.}}
\usepackage{dcolumn}
\usepackage{bm}
\usepackage{float}
\usepackage[Gray,squaren,thinqspace,thinspace]{SIunits} 
\usepackage{amsmath}
\usepackage{ifthen}

\newcommand{\SET}[1]  {\ensuremath{\mathcal{#1}}}

\newcommand{\VEC}[1]  {\ensuremath{\boldsymbol{#1}}}

\begin{document}


\title{Forbidden transitions in neutral and charged current interactions between low--energy neutrinos and Argon}


\author{N.~Van Dessel}
\email{Nils.VanDessel@UGent.be}
\affiliation{Department of Physics and Astronomy,\\
Ghent University,\\ Proeftuinstraat 86,\\ B-9000 Gent, Belgium}
\author{N.~Jachowicz}
\email{Natalie.Jachowicz@UGent.be}
\affiliation{Department of Physics and Astronomy,\\
Ghent University,\\ Proeftuinstraat 86,\\ B-9000 Gent, Belgium}
\author{A.~Nikolakopoulos}
\affiliation{Department of Physics and Astronomy,\\
Ghent University,\\ Proeftuinstraat 86,\\ B-9000 Gent, Belgium}



\date{\today}

\date{\today}

\begin{abstract}
\textbf{Background:} The study of low--energy neutrinos and their interactions with atomic nuclei is crucial to several open problems in physics, including the neutrino mass hierarchy, CP--violation, candidates of Beyond Standard Model physics and supernova dynamics. Examples of experiments include CAPTAIN at SNS as well as DUNE's planned detection program of supernova neutrinos. 

\textbf{Purpose:} We present cross section calculations for charged current and neutral current neutrino--nucleus scattering at low energies, with a focus on ${}^{40}$Ar. We also take a close look at pion decay--at--rest neutrino spectra, which are used in e.g. the SNS experiment at Oakridge.

\textbf{Method and results:} We employ a Hartree Fock + Continuum Random Phase Approximations (HF+CRPA) framework, which allows us to model the responses and include the effects of long--range correlations. It is expected to provide a good framework to calculate forbidden transitions, whose contribution we show to be non--negligible.
 
 \textbf{Conclusions:} Forbidden transitions can be expected to contribute sizeably to the reaction strength at typical low--energy kinematics, such as DAR neutrinos. Modeling and Monte Carlo simulations need to take all due care to account for the influence of their contributions.
\end{abstract}

\maketitle

\section{Introduction}\label{sec:int}

Within the realm of neutrino physics, one area of research that has a lot of potential for exciting new discoveries lies in the study of low-energy neutrinos. The interaction of these with atomic nuclei provides crucial information needed for several open questions in physics. DUNE~\cite{DUNE,Acciarri:2015uup} can, for example, as part of their low--energy (LE) neutrino program, distinguish between the two hierarchy scenarios in the detection of supernova neutrinos as explained in e.g.~\cite{Ankowski:2016lab}, where mention is also made of the prospect to probe Beyond Standard Model (BSM) physics such as sterile neutrinos, neutrino magnetic moments and Non--Standard Interactions (NSI). Furthermore, the measurement of low--energy neutrinos is crucial to supernova research, where detecting the outgoing neutrinos can be used to study the underlying supernova dynamics and the interactions that the neutrinos engage in~\cite{Ankowski:2016lab}. Finally, another avenue worth mentioning lies in the experimental measurement of coherent elastic neutrino-nucleus scattering (CE$\nu$NS)~\cite{1742-6596-606-1-012010,coherent,Akimov:2017ade}.

\begin{figure}
   \centering
   \includegraphics[width=0.95\columnwidth]{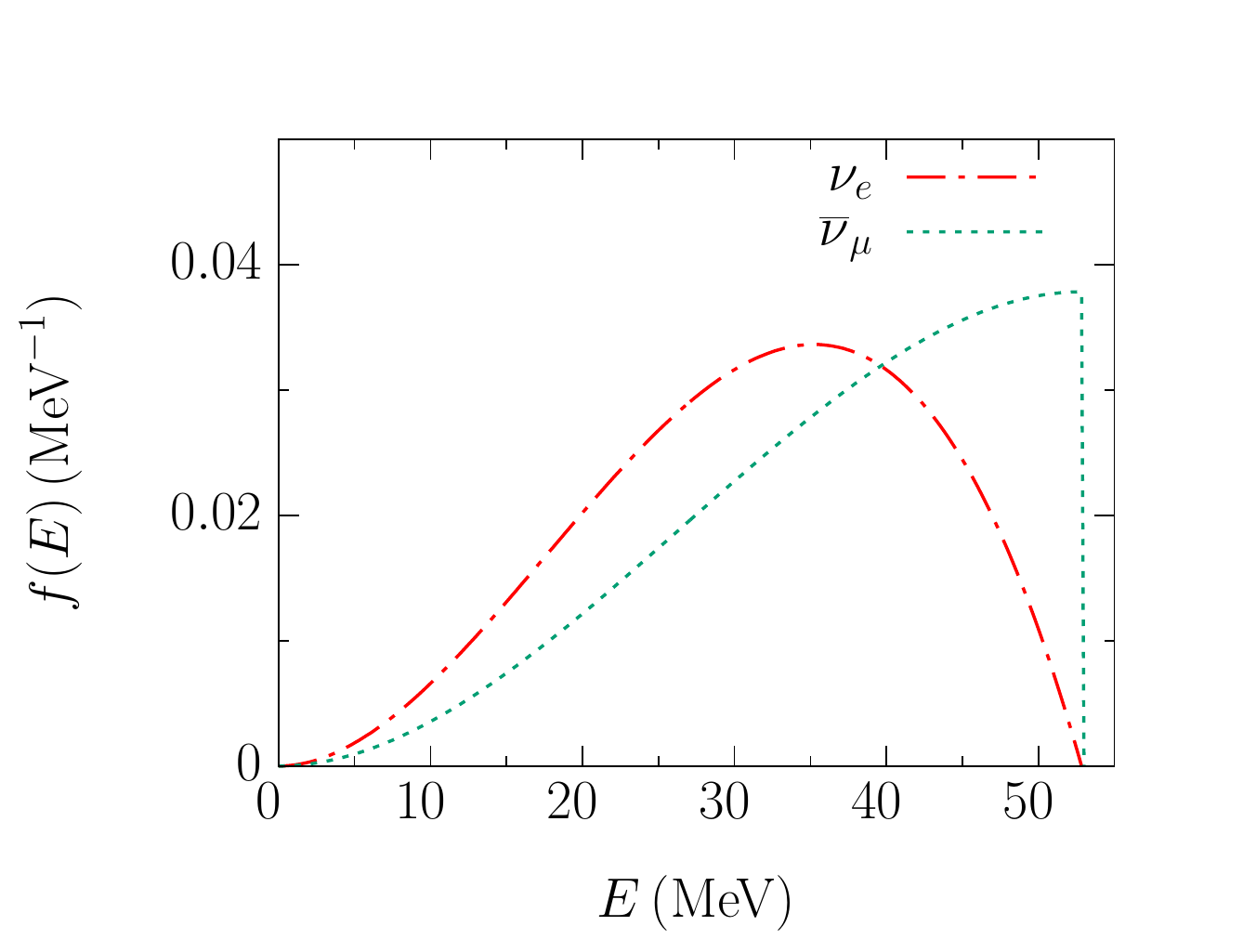}
   \caption{The normalized electron neutrino and muon antineutrino fluxes produced by DAR pions, such as at SNS.}
   \label{fig:michel}
\end{figure} 

Inelastic cross sections induced by low--energy neutrinos (10s of MeVs) have been measured in the past for carbon and iron by the LSND and KARMEN collaborations~\cite{Auerbach:2001hz,Maschuw:1998qh}. Several experiments have been proposed for the near future to measure supernova neutrinos at higher precision, including the aforementioned DUNE as well as e.g.~JUNO~\cite{juno} and Hyper-Kamiokande~\cite{hyperk}. These require significant research and development efforts in order to properly calibrate the detectors that are to be used, including a better knowledge of the cross sections. One necessity is to have access to neutrinos of similar kinematics to those produced in supernovae. The Spallation Neutrino Source at Oakridge (SNS)~\cite{Efremenko:2008an} is an example of such a facility. As a byproduct of its primary purpose of producing neutrons, the SNS also creates neutrinos of several flavors. These are born out of pions decaying at rest (DAR), yielding monoenergetic muon neutrinos at $E_{\nu_\mu} = 29.8$ MeV as well as electron neutrinos and muon antineutrinos, with their normalized flux pictured in Fig.~\ref{fig:michel}, the \mbox{well--known} Michel spectra. Part of the CAPTAIN~\cite{captain} program is to have the CAPTAIN Liquid Argon Time Projection Chamber (LarTPC) detector run at the SNS~\cite{Berns:2013usa}. This will yield crucial information on the neutrino-nucleus cross sections, but also on how to characterize these events in the analyses. A proper analysis of all the produced particles is a necessity to perform accurate calorimetry in SN neutrino experiments. Since it is of relevance to further discussions in this paper, we also make mention of the fact that specialized event generators are required for simulations, since most like NuWro~\cite{nuwro}, NEUT~\cite{Hayato:2009zz} and GENIE~\cite{genie} are tailored towards higher-energy scenarios. DUNE uses MARLEY~\cite{marley}, for example, to model charged current interactions between LE neutrinos and ${}^{40}$Ar.

Parallel to all of these developments lies a need to theoretically study and model cross-sections for charged current (CC) and neutral current (NC) interactions of SN neutrinos with nuclei. Past efforts to that effect include calculations performed in RPA frameworks such as Refs.~\cite{McLaughlin:2004va,Engel:2002hg,Volpe:2001gy,Auerbach:1997ay,SUZUKI2003446,Bandyopadhyay:2016gkv,Cheoun:2011zza}, shell model~\cite{Hayes:1999ew,Suzuki:2006qd,Suzuki:2013wda,Kostensalo:2018kgh} (or hybrid models of the two~\cite{Kolbe:2000np,Kolbe:1999vc,Suzuki:2009zzc,Suzuki:2012ds}) as well as QRPA~\cite{Paar:2007fi,Samana:2008pt,Cheoun:2010pn}, relativistic (R)QRPA~\cite{Paar:2011pz} and local Fermi Gas--based RPA approaches~\cite{Nieves:2004wx,SajjadAthar:2005ke,Singh:1998md}. We also mention previous CRPA results in~\cite{Kolbe:1999au,Jachowicz:2002hz,Jachowicz:2003iz}. Comparisons between shell model, RPA and CRPA results were performed in~\cite{Volpe:2000zn,Kolbe:2003ys}. Phenomenological calculations of cross sections were also provided, in Refs.~\cite{Pourkaviani:1990et,Fukugita:1988hg,Mintz:2002rt}. In this paper, we present cross section calculations for low--energy CC and NC neutrino--nucleus scattering in a HF+CRPA approach, with a focus on ${}^{40}$Ar. 

\section{Model}\label{sec:for}
We will briefly discuss the model used for our calculations. The cross section for inclusive electroweak scattering of neutrinos off atomic nuclei is given by the familiar expression:

\begin{figure}
  \centering
  \includegraphics[width=0.85\columnwidth]{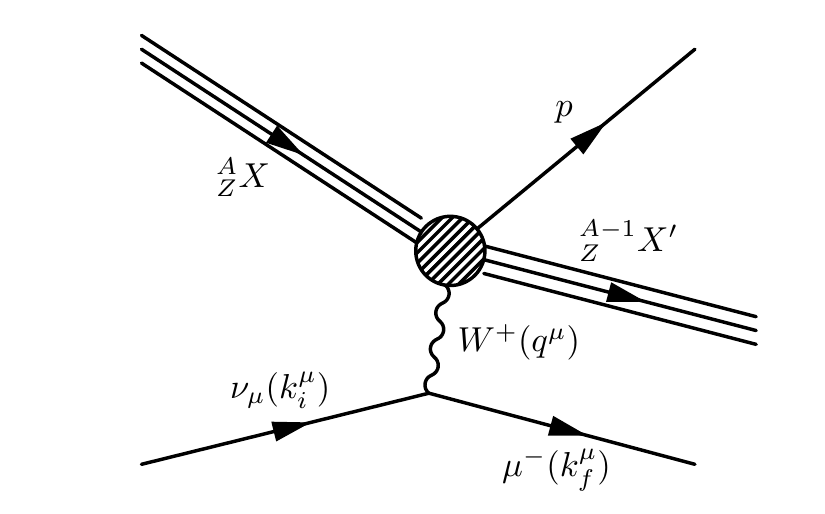}
  \caption{Diagrammatical representation of neutrino--nucleus scattering, pictured here for the case of a CC interaction.}
  \label{fig:diagram}
\end{figure}

\begin{equation}\label{eq:xsec}
\begin{aligned}
\frac{\mathrm{d}\sigma}{\mathrm{d}T_f\mathrm{d}\Omega_f} =& \sigma_X E_f k_f \zeta^2(Z',E_f) \\
&\times \left( v_{CC} W_{CC} + v_{CL} W_{CL} + v_{LL} W_{LL} \right.  \\
& + \left. v_{T} W_{T} \pm v_{T'} W_{T'} \right),
\end{aligned}
\end{equation}

where $\sigma_X$ is $\left(\frac{G_F \cos{\theta_c}}{2\pi} \right)^2$ for CC scattering and $\left(\frac{G_F}{2\pi} \right)^2$ for NC. In turn, $G_F$ is the Fermi constant that encodes the strength of the weak interaction, while $\cos{\theta_c}$ is the Cabibbo angle. The factor $\zeta^2(Z',E_f)$ is related to the Coulomb interaction between the escaping lepton and the residual nucleus. In the case of NC, it is simply 1. We will come back to the specifics of this factor further on. The cross section also differs for neutrinos and antineutrinos due to the V--A nature of the electroweak interaction, as shown by the $\pm$--sign in the transverse interference term. The $v$--factors are the leptonic functions, while the $W$--factors are the nuclear response functions, which contain all the nuclear information on the reaction to the electroweak probe. They are a function of the transition amplitude between the initial $| \Phi_\textrm{0} \rangle$ and final $| \Phi_\textrm{f} \rangle$ state:

\begin{equation}\label{eq:current}
\SET{J}_\lambda^{nucl}(\omega,\VEC{q}) = \langle \Phi_\textrm{f} | \hat{J}_\lambda(\VEC{q}) | \Phi_\textrm{0} \rangle .
\end{equation}

This nuclear current is the Fourier transform of the current in coordinate space:
\begin{equation}
\hat{J}_\lambda(\VEC{q}) = \int \mathrm{d}\VEC{x} e^{i\VEC{x}\cdot\VEC{q}} \hat{J}_\lambda(\VEC{x}),
\end{equation}
where $\omega$ and $\VEC{q}$ are the energy and momentum transfer, respectively. One performs a multipole expansion of the nuclear current, which can be used to exploit angular momentum algebra. If one chooses the z--direction along the momentum transfer $\VEC{q}$:

\begin{equation}
\begin{aligned}
J_0(\VEC{q}) =& \sqrt{4\pi} \sum_{J=0}^{+\infty} i^J \sqrt{2J+1}\hat{M}_{J,0}^{C}(q) \\
J_3(\VEC{q}) =& -\sqrt{4\pi} \sum_{J=0}^{+\infty} i^J \sqrt{2J+1}\hat{L}_{J,0}^{L}(q) \\
J_{\pm}(\VEC{q}) =& -\sqrt{2\pi} \sum_{J=1}^{+\infty} i^J \sqrt{2J+1}\left[\hat{T}_{J,\pm 1}^{E}(q) \pm \hat{T}_{J,\pm 1}^{M}(q) \right].
\end{aligned}
\end{equation}

Expressions for the nuclear one--body current and derived Coulomb ($\hat{M}_{J,L}^{C}$), longitudinal ($\hat{L}_{J,L}^{L}$), electric ($\hat{T}_{J,L}^{E}$) and magnetic ($\hat{T}_{J,L}^{M}$) multipole operators can be found in e.g.~\cite{walecka2004theoretical}. Worth mentioning at this stage in the context of low--energy neutrinos, is the 'allowed approximation'. Under this approximation, valid at low energies and typically used in beta--decay calculations, it is customary to employ both the long--wavelength limit $q \rightarrow 0$, and to assume that one is dealing with slow nucleons $p_N/m_N \rightarrow 0$. Under these assumptions, one can show that the only terms surviving in the multipole decomposition of the nuclear current for CC interactions (similar considerations hold for the NC current) are the following:
\begin{equation}
\begin{aligned}
\hat{M}_{J,0}^{C} =& \frac{1}{\sqrt{4}}F_1 \tau_{\pm}(i),\\
\hat{T}_{1,m}^{E} =& \sqrt{2} \hat{L}_{1,m}^{L} = \frac{i}{\sqrt{6}}G_A\sum_{i=1}^{A}\tau_{\pm}(i)\sigma_{1,m}(i),
\end{aligned}
\end{equation}
where $F_1$ and $G_A$ are the Fermi and axial form factors, respectively. These are the well--known Fermi and Gamow--Teller transition operators. At sufficiently low energies, they provide an adequate description of the nucleus' response to an electroweak probe. Because they are responsible for the largest part of the reaction strength, the transitions they give rise to are known as 'allowed', with the $J^P$ quantum numbers of the operators $0^+$ and $1^+$, respectively. Conversely, higher--order transitions are often referred to as 'forbidden' transitions. As we will show, the latter are still important at the energies considered in this context. In this paper, the nuclear responses are calculated in a Continuum Random Phase Approximation (CRPA), where long--range correlations and collective excitations of the nucleus are taken into account. We mention that in these calculations, we use the free-nucleon value for the axial coupling of $g_A = 1.27$, whereas some models use an effective quenched value of around $g_A =  1.00$. This issue is discussed in Ref.~\cite{Pastore:2017uwc}, where it is argued that the need for such a quenching may appear as a consequence of an absence of nuclear correlations or an insufficiently large model space in the modeling.

Within the HF+CRPA approach, excited states of the nucleus are described as coherent superpositions of particle--hole and hole--particle excitations out of the correlated ground state:

\begin{equation}
| \Psi^c_{RPA} \rangle = \sum_{c'} \left\lbrace X_{c,c'} |p'h'^{-1}\rangle - Y_{c,c'} |h'p'^{-1}\rangle \right\rbrace,
\end{equation}
where c contains all the quantum numbers to unambiguously label an excitation channel. The local RPA polarization propagator, containing all the information on excited states, is implicitly defined as
\begin{equation}
\begin{aligned}
&\Pi^{RPA}(x_1,x_2,E_{exc}) = \Pi^{(0)}(x_1,x_2,E_{exc}) \\ 
&+ \frac{1}{\hbar} \int \mathrm{d}x \int \mathrm{d}x' \left[ \Pi^{(0)}(x_1,x,E_{exc}) \right. \\
& \left. \times \tilde{V}(x,x') \Pi^{RPA}(x',x_2,E_{exc}) \right],
\end{aligned}
\end{equation}
with $ \tilde{V}(x,x')$ the antisymmetrized residual interaction, $E_{exc}$ the excitation energy of the nucleus and $x$ a shorthand for spatial, spin and isospin coordinates. The residual interaction used is the same Skyrme force that is employed to calculate the single--particle wave functions, so that the scheme is self--consistent. Partially filled shells such as those present in $^{40}$Ar are dealt with by including occupation probabilities in the transition amplitudes:
\begin{equation}
\langle ph^{-1} | \hat{O} | \Phi_\textrm{0} \rangle \rightarrow v_h \langle ph^{-1} | \hat{O} | \Phi_\textrm{0} \rangle,
\end{equation}
with $\hat{O}$ representing a general one--body operator and $v_h^2$ being the occupation probability of the shell $h$. Further details and previous applications of this model can be found in Refs.~\cite{Ryckebusch:1988aa, Ryckebusch:1989nn, Jachowicz:1998fn, Jachowicz:2002rr,Jachowicz:2002hz,Jachowicz:2004we, Jachowicz:2006xx, Jachowicz:2008kx,Pandey:2014tza, Pandey:2016jju,VanDessel:2017ery,Nikolakopoulos:2018sbo}. While our calculations predominantly pertain to describing reactions in the Giant Resonance (GR) region due to the low neutrino energies, the HF+CRPA model described above is also successful at describing the quasielastic region at higher energies, such as those relevant to baseline experiments~\cite{Pandey:2013cca,Pandey:2014tza,Pandey:2015gta,Pandey:2016jju,VanDessel:2017ery}. \\

\begin{figure}
   \centering
   \includegraphics[width=0.95\columnwidth]{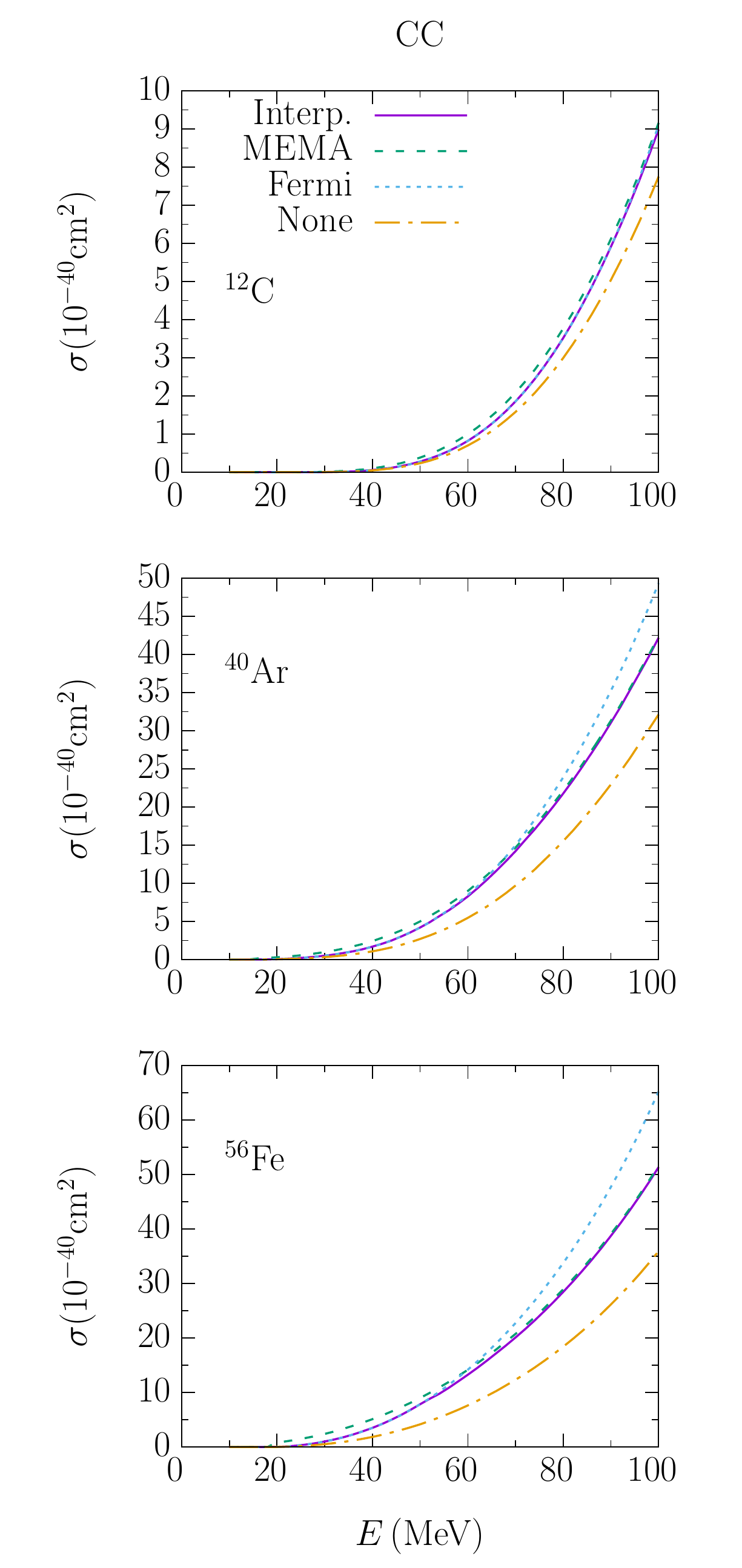}
   \caption{Total charged--current cross sections, shown for various nuclei and for various Coulomb schemes, as a function of incoming neutrino energy $E$.}
   \label{fig:fecoulombcomp}
\end{figure}

Finally, we turn our attention to the topic of Coulomb interactions between the outgoing lepton and the residual nucleus in charged--current interactions. At low energies, this is expected to have a strong effect on the cross section due to the low momentum of the charged outgoing lepton. In our discussion we will consider two effective schemes to account for the Coulomb distortion. At low outgoing lepton energies, of the order of magnitude applicable to e.g. beta decay, one can make use of the Fermi function~\cite{Engel:1997fy}:
\begin{equation}
\begin{aligned}
\zeta^2(Z',E_f) =& 2(1+\gamma_0)(2p_f R)^{-2(1-\gamma_0)} \\
\times& \frac{|\Gamma(\gamma_0+i\eta)|^2}{(\Gamma(2\gamma_0+1))^2}e^{\pi \eta},
\end{aligned}
\end{equation}
where $R\approx 1.2 A^{1/3} \mathrm{fm}$ is the nuclear radius, $\gamma_0=\sqrt{1-(\alpha Z')^2}$, $E_f$ is the outgoing lepton's energy, $p_f$  outgoing momentum and $\eta=\pm \frac{\alpha Z' c}{v}$. with $+$ and $-$ for neutrinos and antineutrinos respectively. Similarly, the final nuclear charge $Z'$ is equal to $Z+1$ or $Z-1$ for $\nu$/$\bar{\nu}$, respectively. This approximation assumes that the outgoing leptons only contribute sizeably as an s--wave to the reaction strength, and is therefore not applicable at higher outgoing lepton energies~\cite{Engel:1997fy}. Therefore, in these regimes, we will consider a different scheme, the modified effective momentum approximation (MEMA), detailed in Ref.~\cite{Engel:1997fy}, where the energy and momentum of the final lepton is shifted to an effective value by the Coulomb energy in the center of the nucleus:
\begin{equation}
E_{eff} = E_f - V_c(0) = E \pm \frac{3}{2}\frac{Z' \alpha \hbar c }{R},
\end{equation}
which introduces a factor in the cross section that accounts for the change in phase space:
\begin{equation}
\zeta^2(Z',E_f) = \frac{E_{eff} k_{eff} }{E_f k_f},
\end{equation}
and induces a shift in the momentum transfer $q \rightarrow q_{eff}$ in the calculation of the amplitudes in Eq.~(\ref{eq:current}). In practice, we will interpolate between the two schemes. This approach consists of taking for each value of $\omega$ in the differential cross section the smallest or highest value of the two, respectively for neutrinos and antineutrinos where the effect of the Coulomb interaction is to increase resp. decrease both the differential and the integrated cross section~\cite{Engel:1997fy}. In short: we take the value of $\zeta^2(Z',E_f)$ that is closest to unity. The effect of this is shown in Fig.~\ref{fig:fecoulombcomp}. At low incoming neutrino energies, the interpolation scheme matches that of the Fermi function, whilst at higher energies, where reaction strength contains increasingly more contributions from events with high momentum of the final lepton, the scheme matches that of the MEMA. We show this for several nuclei. For ${}^{12}$C, it is clear that the Fermi function is a good approximation over the whole energy range, while for ${}^{40}$Ar and ${}^{56}$Fe, the Fermi function and MEMA schemes cross between 60 and 70 MeV. In general, the heavier the nucleus, the more limited the Fermi function will be in its area of applicability. A similar comparison of Coulomb schemes was performed in Ref.~\cite{SajjadAthar:2005ke}.

\section{Results}\label{sec:res}

\begin{figure}
   \centering
   \includegraphics[width=0.95\columnwidth]{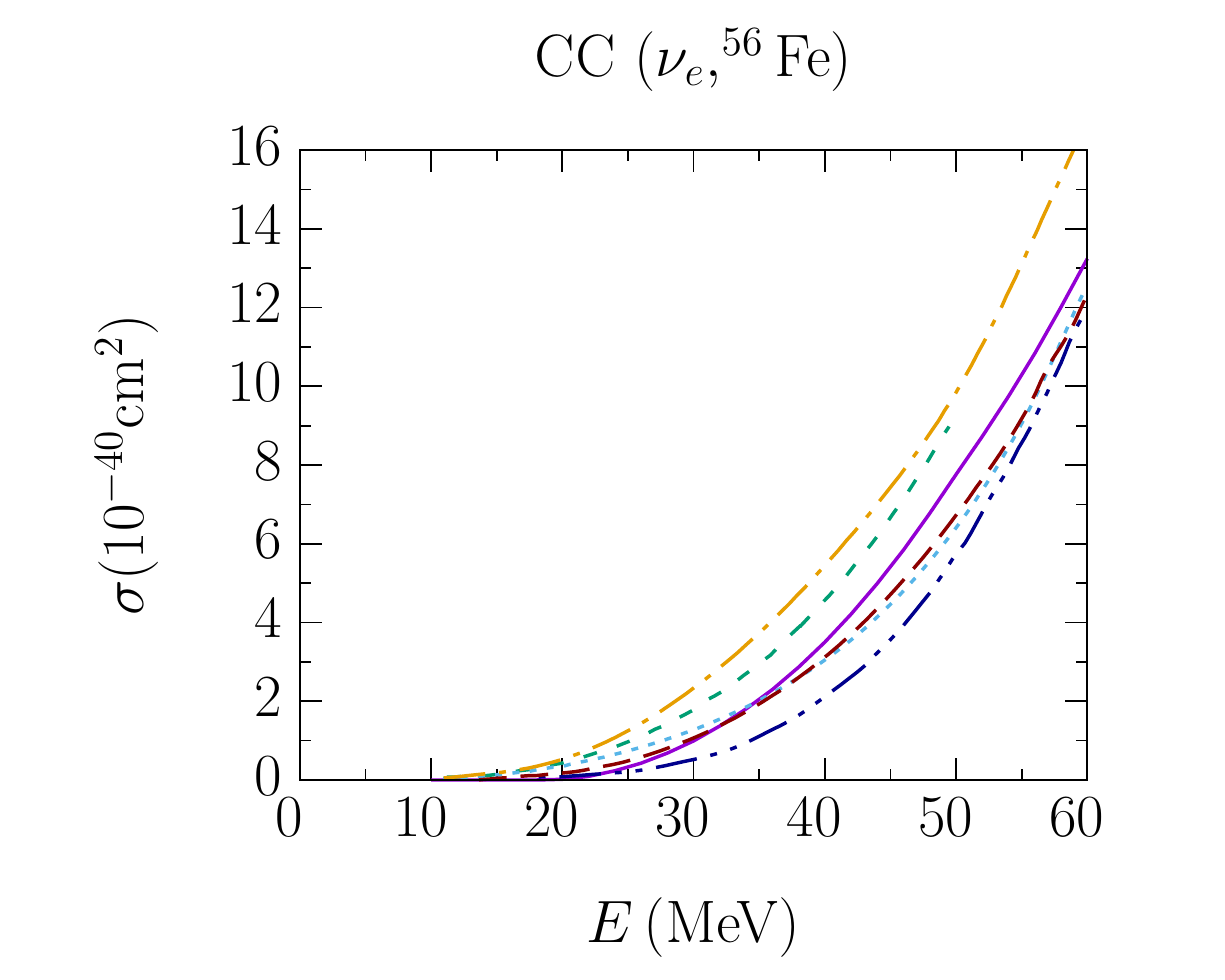}
   \caption{The total charged--current ($\nu_e$,${}^{56}$Fe) cross section, shown for various models. The solid curve represents the HF+CRPA result, the short--dashed curve is taken from~\protect\cite{SajjadAthar:2005ke}, the dotted curve from~\protect\cite{Bandyopadhyay:2016gkv}, the dot--dashed curve from~\protect\cite{Lazauskas:2007bs}, the double dot--dashed curve from~\protect\cite{Paar:2007fi} and finally the long--dashed curve from~~\protect\cite{Samana:2008pt}}
   \label{fig:femodelcomp}
\end{figure}

Before we cover the results for neutrino--argon calculations, we wish to compare our model predictions with other recent studies at these energies for various nuclei. In Fig.~\ref{fig:femodelcomp} we confront the HF+CRPA results for ${}^{56}$Fe to those of other models. The comparison is fine, except that the strength lies comparatively low for lower energies. This is because of the fact that, within the CRPA, discrete excitations are not taken into account. We can also take a look at the cross section predictions for DAR kinematics, by folding this cross section with the appropriate neutrino spectrum. We compare this to the KARMEN experimental results and various other theoretical models in Table~\ref{table:fedarcomp}. Our result is compatible with experimental results, but has a lower predicted value than most other theoretical models. This is due to the fact that, once again, discrete excitations are not included in the reaction strength. Finally, we compare the $\mathrm{B(GT^-)}$ strength with other models discussed in \cite{Paar:2011pz}. Firstly, this strength, as a function of the nuclear excitation $E_x$ is shown in Fig. ~\ref{fig:fegt}. The HF+CRPA results lack the discrete excitation strength, but reproduces the large peak of the QRPA results relatively well. Secondly, we similarly compare the total strength in Table~\ref{table:fegtcomp}. HF+CRPA underpredicts the other models, likely once again due to the absence of discrete excitations, but still falls within the broad experimental error interval.\\
   {\renewcommand{\arraystretch}{1.2}
\begin{table}
   \centering
\begin{tabular}{|l|l|}
  \hline
   & $\langle \sigma_{DAR} \rangle$ ($10^{-42}\mathrm{cm}^2$) \\
   \hline
   HF+CRPA & 212.9\\
      \hline
  G--Matrix QRPA~\cite{Cheoun:2010pn} & 173.5 \\
  \hline
  Phenomenological~\cite{Mintz:2002rt} & 214   \\
  \hline
  Hybrid~\cite{Kolbe:2000np} & 240\\
  \hline
  Hybrid~\cite{Paar:2011pz} & 259\\
  \hline
  RHB+RQRPA~\cite{Paar:2011pz} & 263\\
  \hline 
  LFG+RPA~\cite{SajjadAthar:2005ke} & 277\\
  \hline
  QRPA~\cite{Lazauskas:2007bs} & 352\\
  \hline 
  Exp. (KARMEN)~\cite{Kolbe:1999vc} & $256\pm108\pm43$ \\
  \hline
\end{tabular}
\caption{The total charged--current ($\nu_e$,${}^{56}$Fe) cross section value, folded with a DAR electron neutrino spectrum, tabulated for various models.}
\label{table:fedarcomp}
\end{table} 
\begin{figure}
   \centering
   \includegraphics[width=0.95\columnwidth]{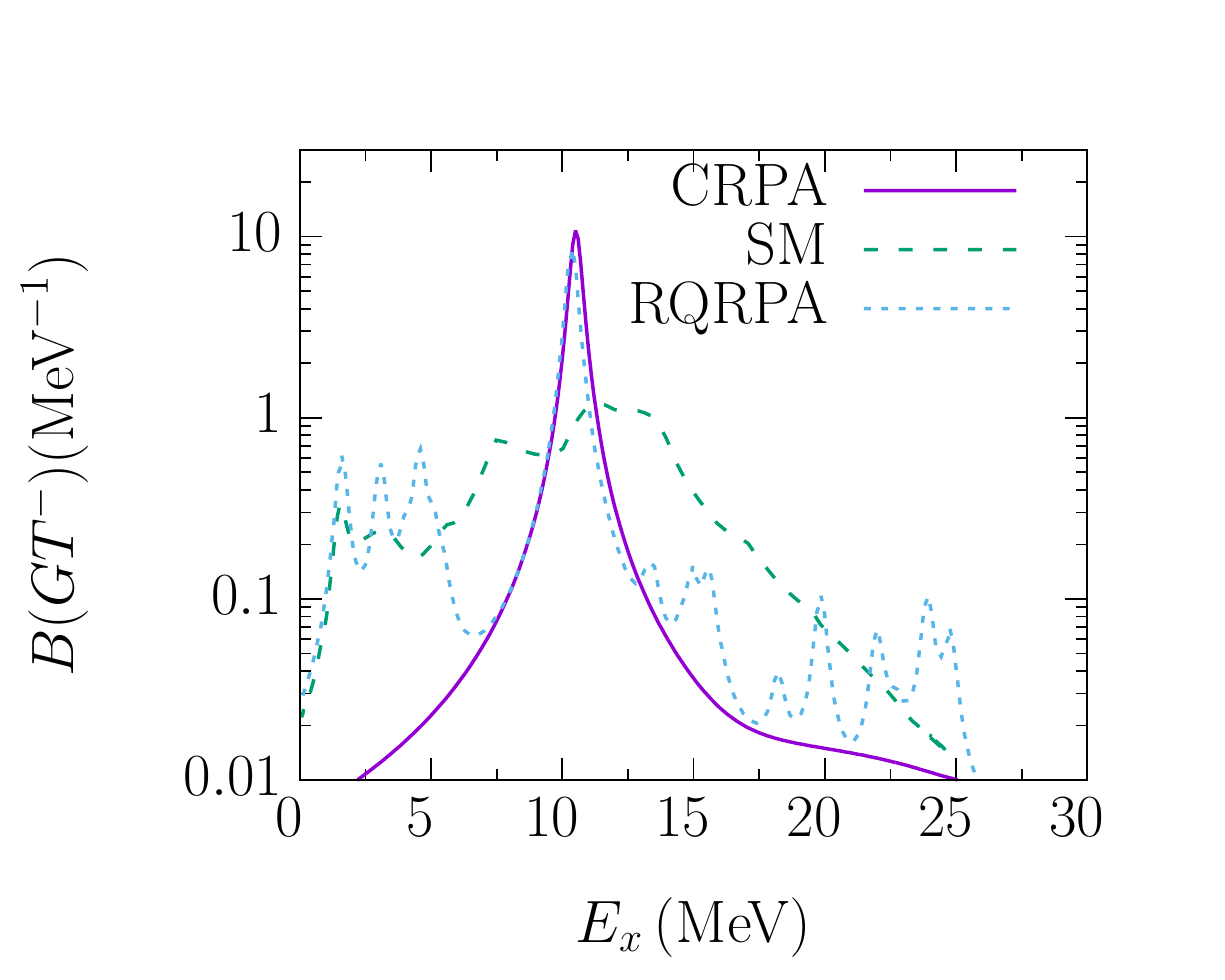}
   \caption{The ${}^{56}$Fe Gamow--Teller strength compared with Shell--Model (long dashed) and QRPA (short dashed) results taken from \cite{Paar:2011pz}. All results folded with a lorentzian of width 0.5 MeV.}
   \label{fig:fegt}
\end{figure} 
  {\renewcommand{\arraystretch}{1.2}
\begin{table}
   \centering
\begin{tabular}{|l|l|}
  \hline
   & $\mathrm{B(GT^-)}$ \\
   \hline
   HF+CRPA & 8.8 \\
      \hline
  GXPF1J  & 9.5 \\
  \hline
  DD--ME2 & 11.3 \\
  \hline
  SGII & 12.3 \\
  \hline
  SLy5 & 14.0 \\
  \hline
  Exp.   & 9.9 $\pm$ 2.5 \\
  \hline 
\end{tabular}
\caption{The total ${}^{56}$Fe $\mathrm{B(GT^-)}$ strength, tabulated for various models from Ref.~\cite{Paar:2011pz}.}
\label{table:fegtcomp}
\end{table} 
\begin{figure}
   \centering
   \includegraphics[width=0.95\columnwidth]{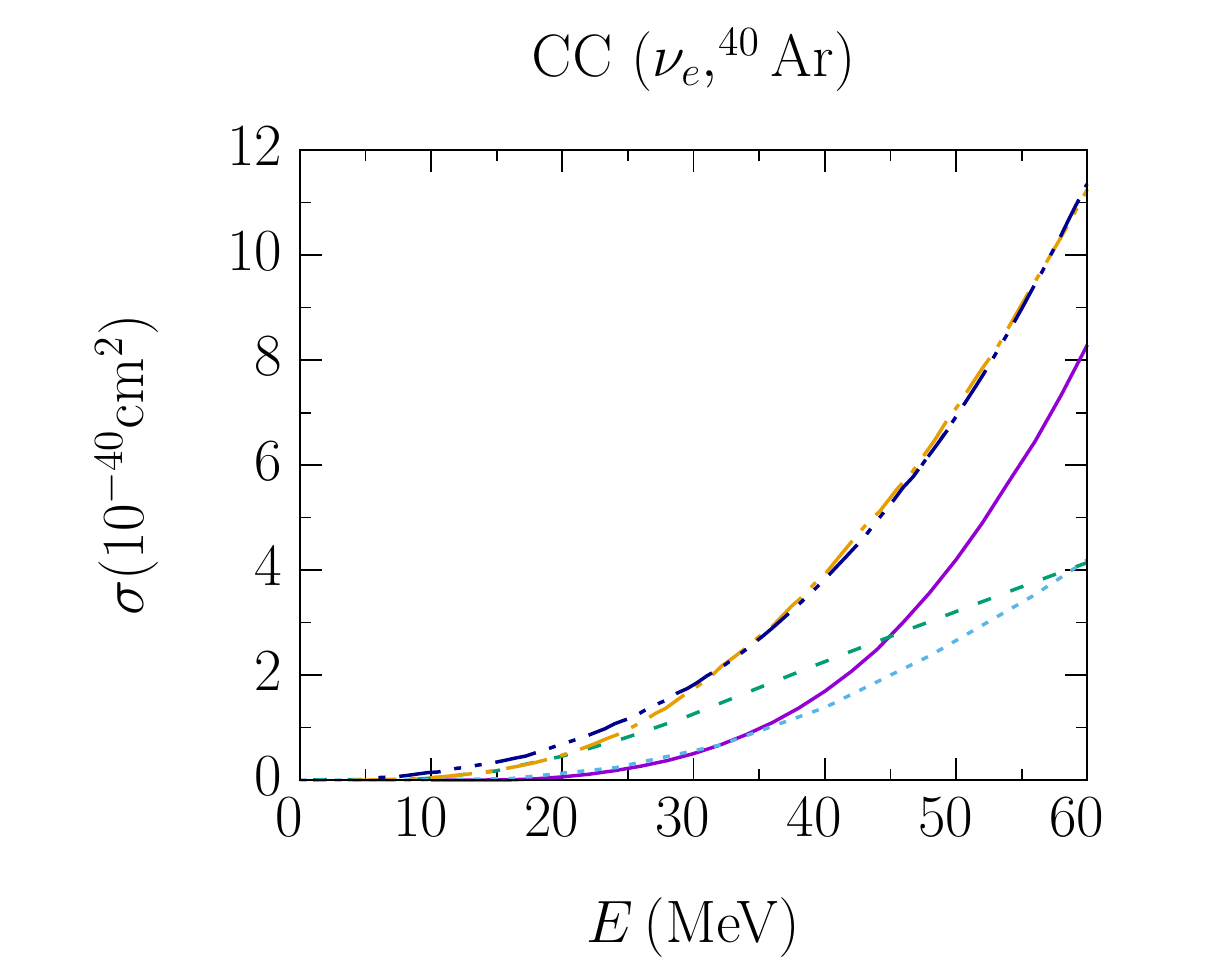}
   \caption{The total charged--current ($\nu_e$,${}^{40}$Ar) cross section, shown for various models. The solid curve represents the HF+CRPA result, the dashed line is the shell model result taken from~\protect\cite{Kostensalo:2018kgh}, the dotted line the QRPA result from~\protect\cite{Cheoun:2011zza}, the dash--dotted line and dash--double dotted line (which overlap to a degree) are from the FG--based RPA result from~\protect\cite{SajjadAthar:2005ke} (using MEMA) and MARLEY~\protect\cite{Gardiner2017}, respectively. }
   \label{fig:armodelcomp}
\end{figure} 
In Fig.~\ref{fig:armodelcomp} we compare with other model predictions for argon. A similar comparison holds. The results from MARLEY match surprisingly well with those from~\cite{SajjadAthar:2005ke}, in spite of using a different, Fermi Gas--based RPA model. The other two models predict lower strength at higher energies. For Ref.~\cite{Kostensalo:2018kgh} in particular, using a shell model, it is argued that this is due to a limited valence space for excited states at higher energies.

\begin{figure}
   \centering
   \includegraphics[width=0.95\columnwidth]{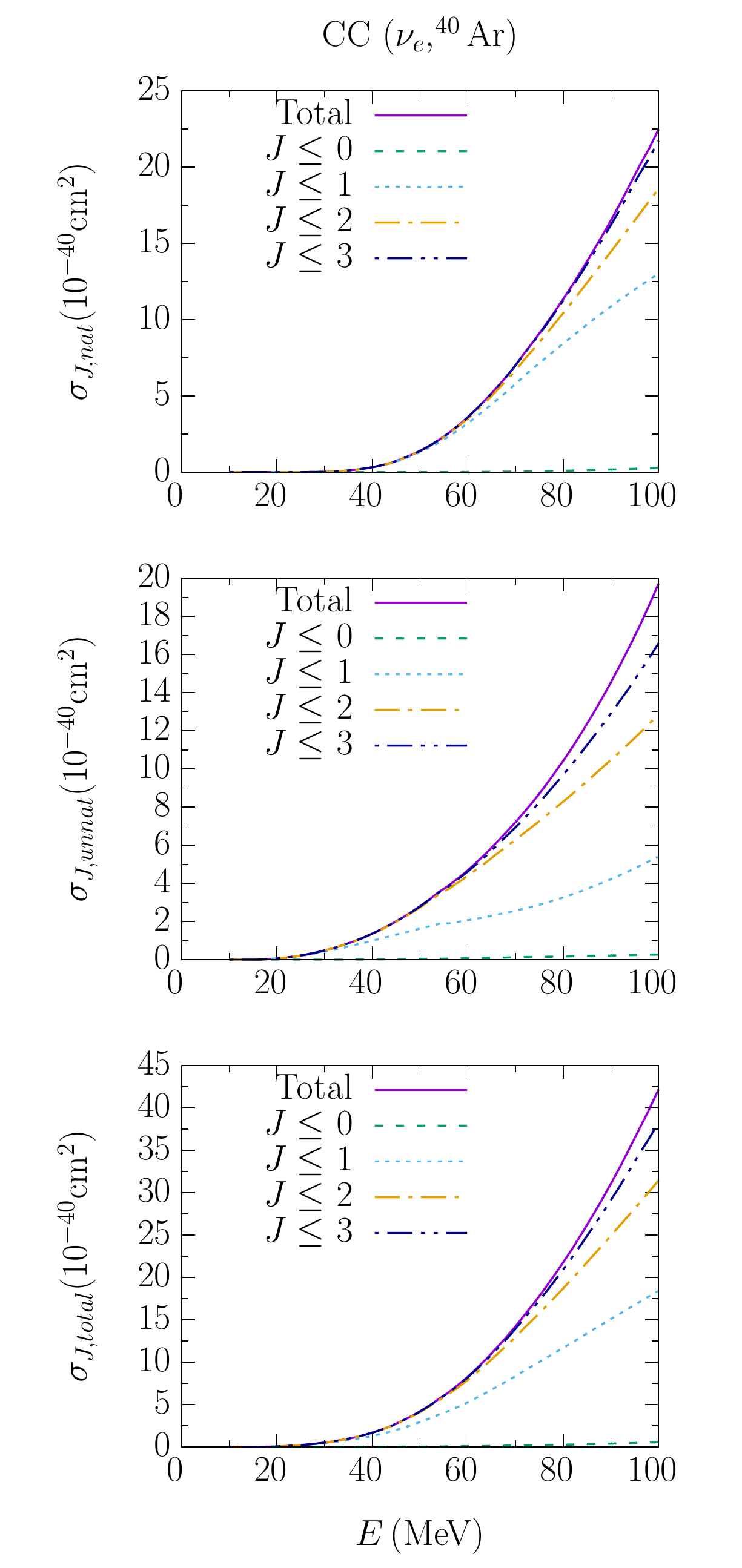}
   \caption{The total charged--current ($\nu_e$,${}^{40}$Ar) cross section, with cumulative reaction strength induced by multipole operators of quantum numbers 0 to J, for natural, unnatural and all transitions.}
   \label{fig:argtotcc}
\end{figure} 

\begin{figure}
   \centering
   \includegraphics[width=0.95\columnwidth]{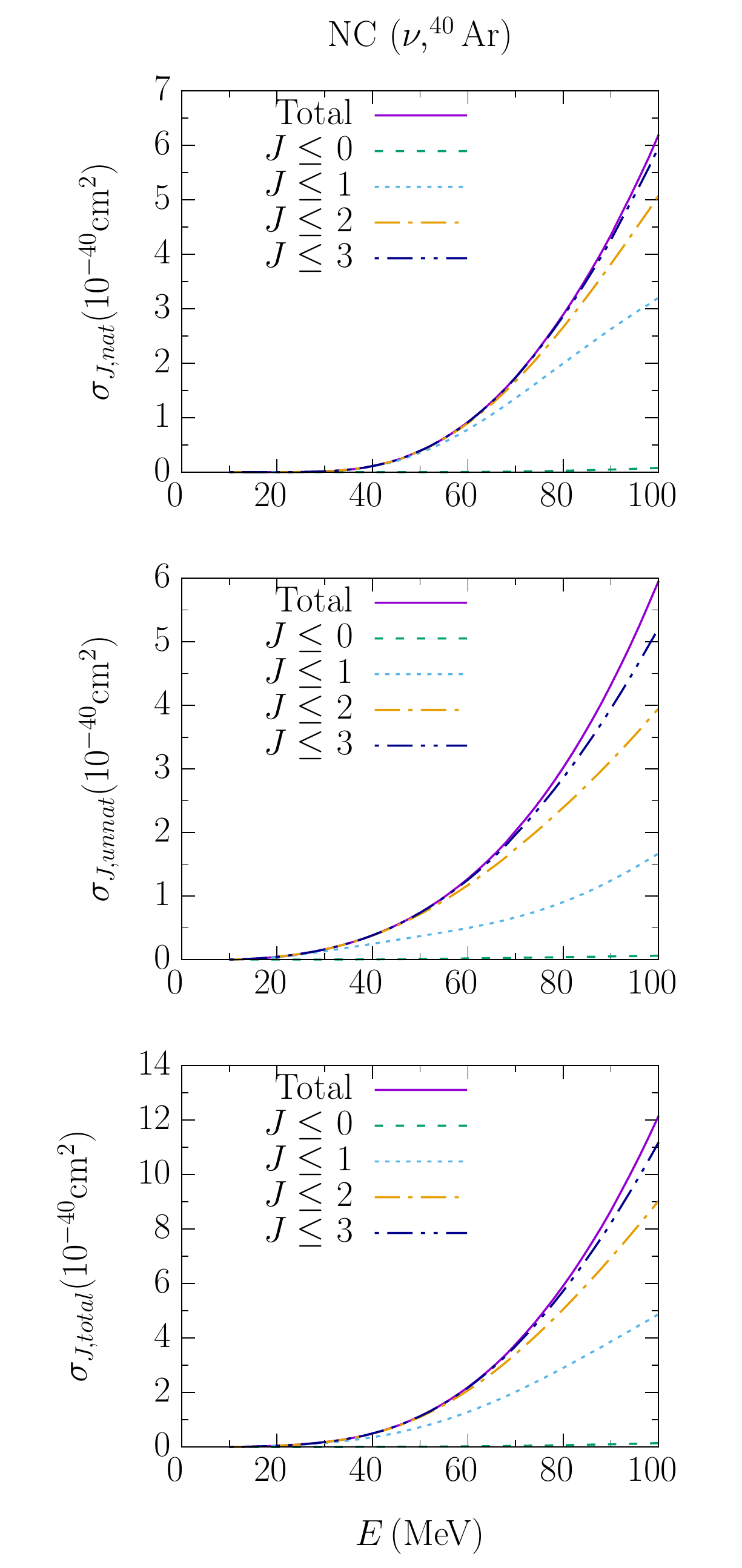}
   \caption{The total neutral--current ($\nu$,${}^{40}$Ar) cross section, with cumulative reaction strength induced by multipole operators of quantum numbers 0 to J, for natural, unnatural and all transitions.}
   \label{fig:argtotnc}
\end{figure} 

\begin{figure}
   \centering
   \includegraphics[width=0.95\columnwidth]{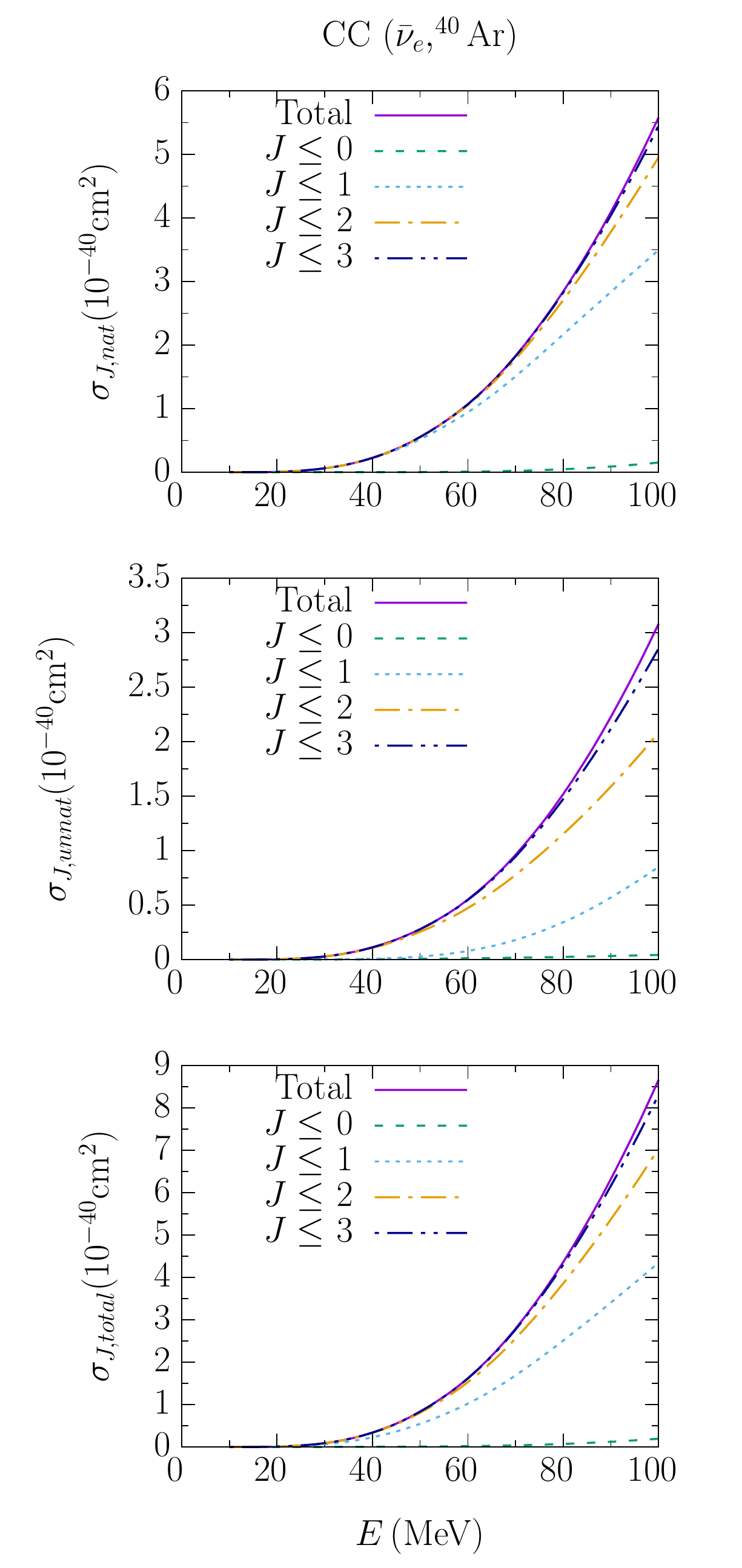}
   \caption{The total charged--current ($\bar{\nu}_e$,${}^{40}$Ar) cross section, with cumulative reaction strength induced by multipole operators of quantum numbers 0 to J, for natural, unnatural and all transitions.}
   \label{fig:argtotccanti}
\end{figure} 

\begin{figure}
   \centering
   \includegraphics[width=0.95\columnwidth]{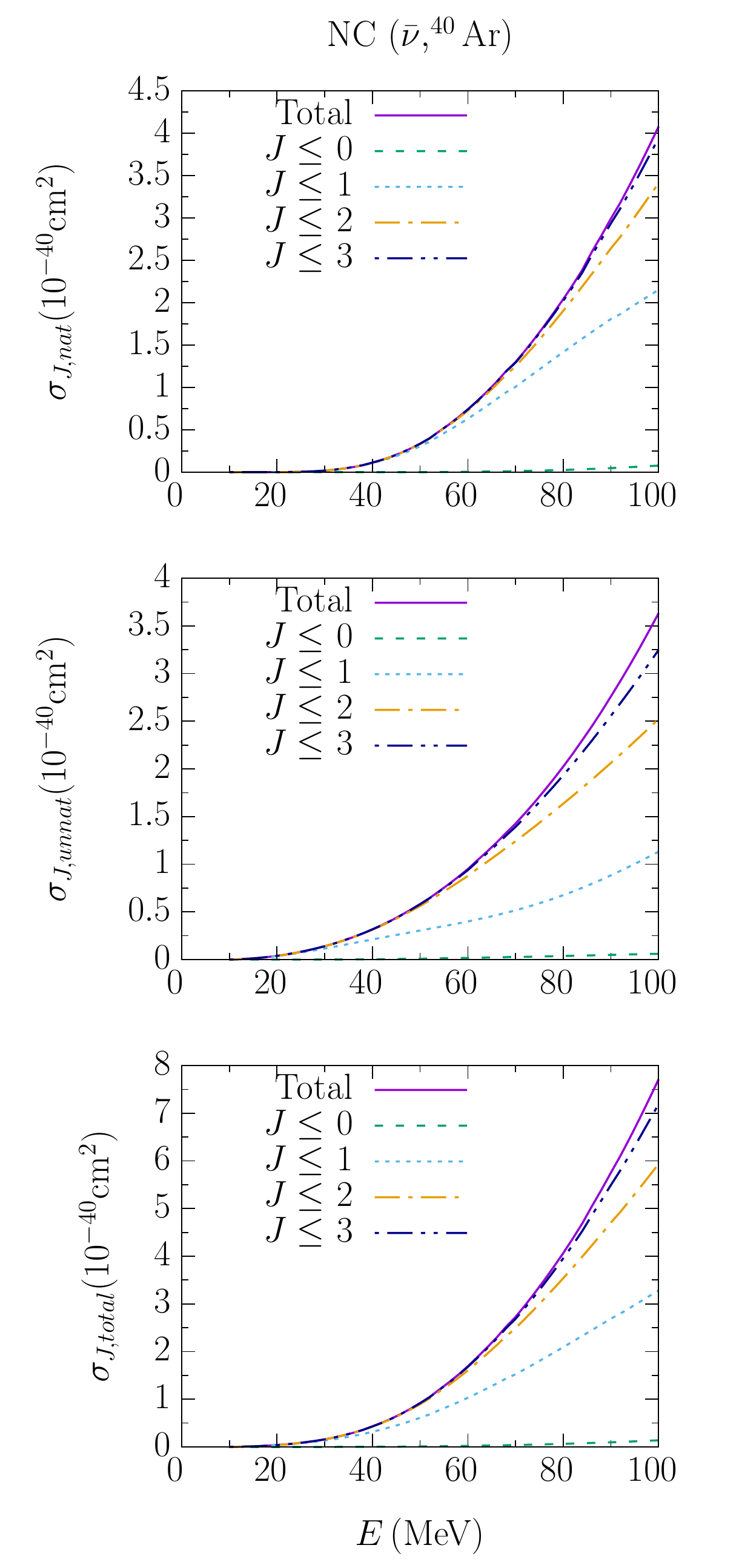}
   \caption{The total neutral--current ($\bar{\nu}$,${}^{40}$Ar) cross section, with cumulative reaction strength induced by multipole operators of quantum numbers 0 to J, for natural, unnatural and all transitions.}
   \label{fig:argtotncanti}
\end{figure}

\begin{figure}
   \centering
   \includegraphics[width=0.95\columnwidth]{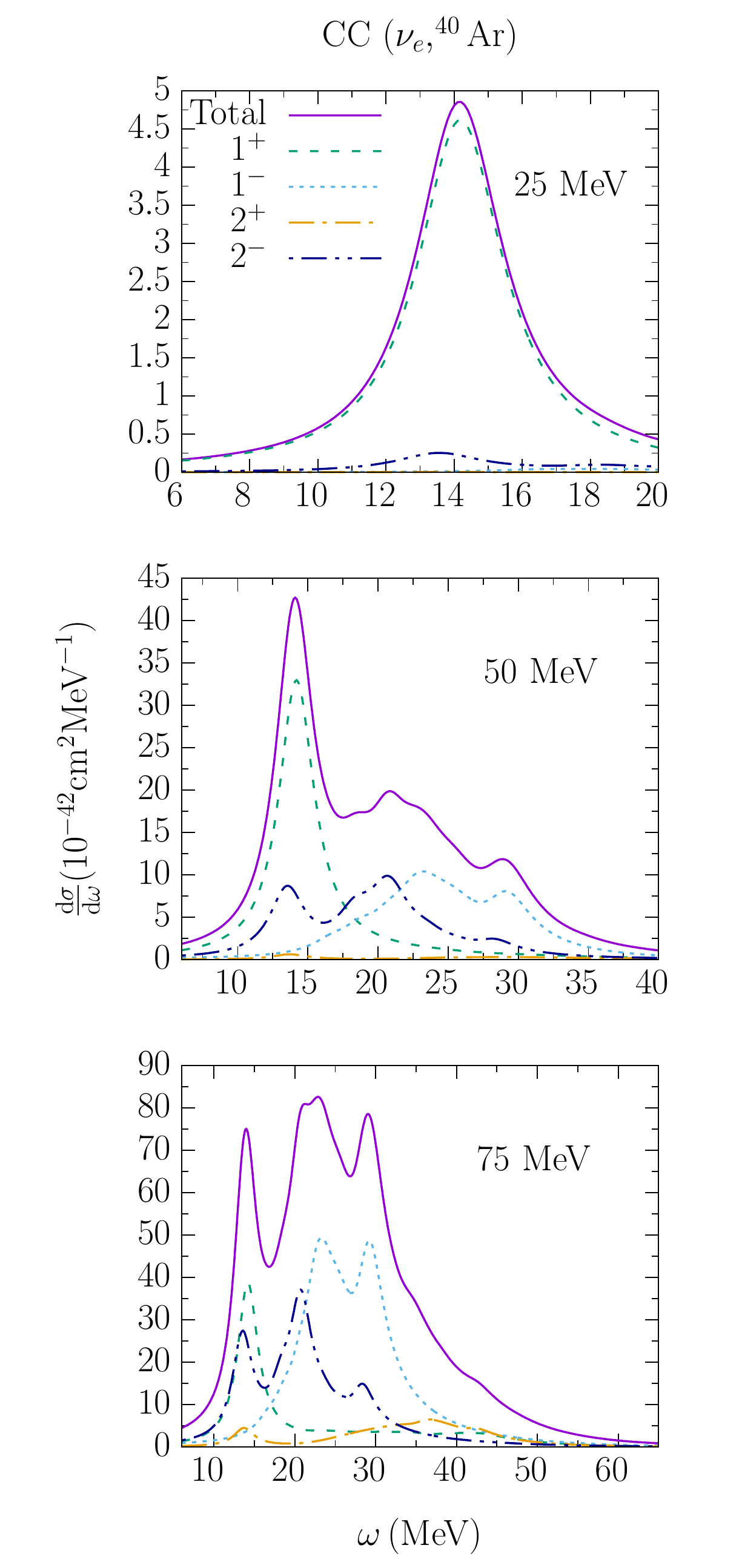}
   \caption{Contributions of different multipole transitions to charged--current ($\nu_e,{}^{40}$Ar) reactions for 25, 50 and 75 MeV incoming neutrino energies. The total includes multipoles up to and including $J=7$. Differential cross sections are folded with a Lorentzian of width 3 MeV in order to account for the finite width of the resonances.}
   \label{fig:singlediffarg}
\end{figure}

\begin{figure}
   \centering
   \includegraphics[width=0.95\columnwidth]{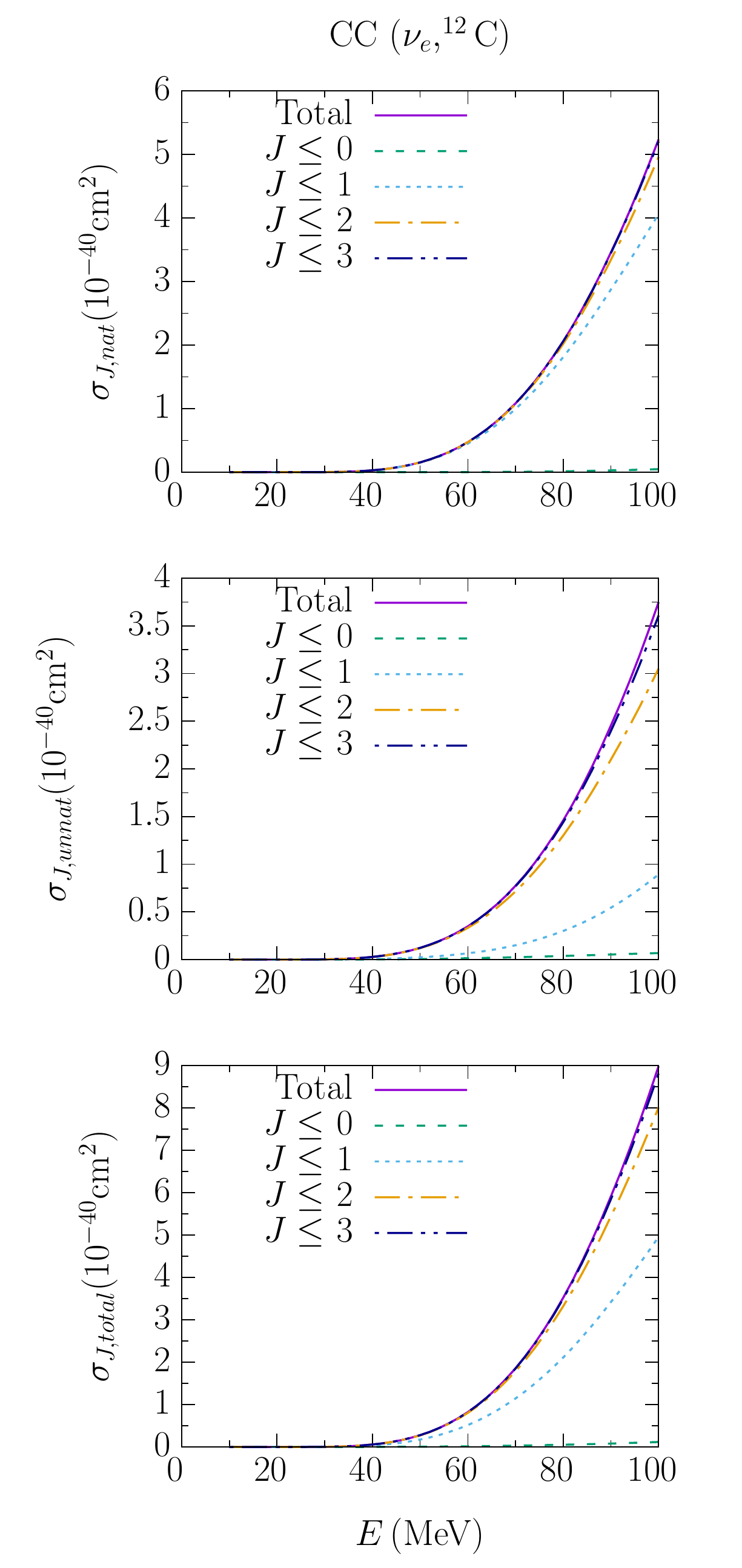}
   \caption{The total charged--current ($\nu_e$,${}^{12}$C) cross section, with cumulative reaction strength induced by multipole operators of quantum numbers 0 to J, for natural, unnatural and all transitions.}
   \label{fig:ctotcc}
\end{figure} 

\begin{figure}
   \centering
   \includegraphics[width=0.95\columnwidth]{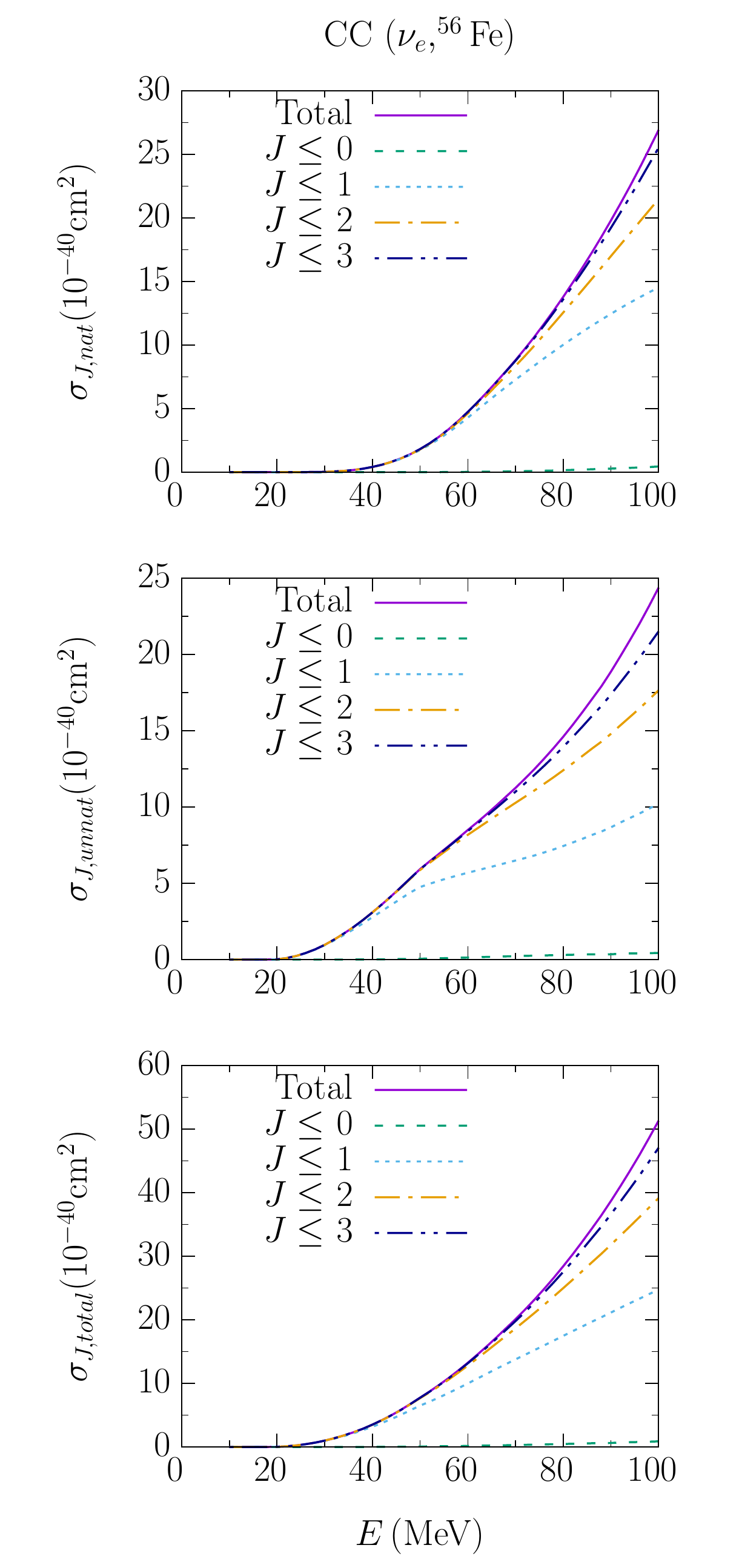}
   \caption{The total charged--current ($\nu_e$,${}^{56}$Fe) cross section, with cumulative reaction strength induced by multipole operators of quantum numbers 0 to J, for natural, unnatural and all transitions.}
   \label{fig:fetotcc}
\end{figure} 

\begin{figure}
   \centering
   \includegraphics[width=0.95\columnwidth]{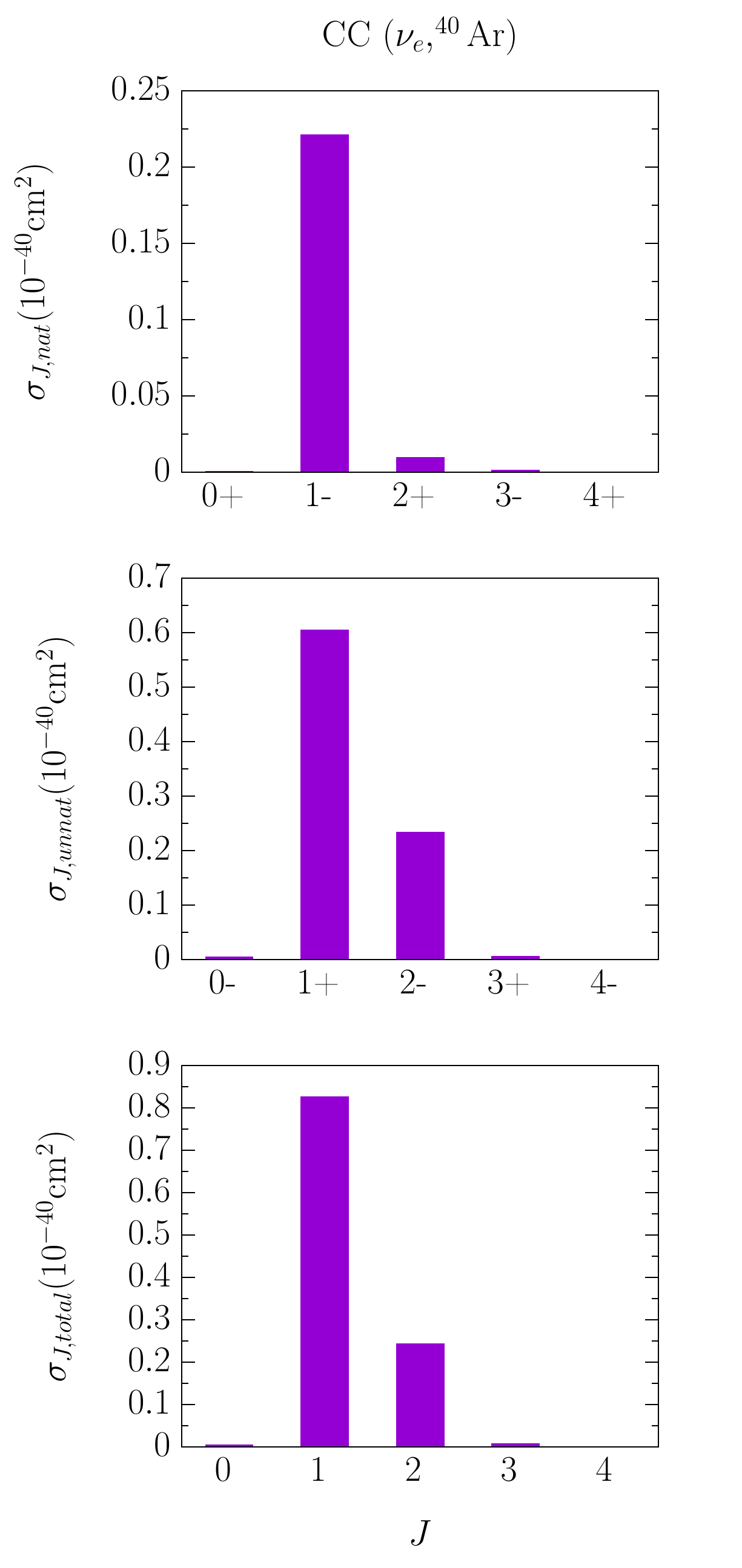}
   \caption{Contributions of different multipole transitions to charged--current reactions for a DAR $\nu_e$ spectrum.}
   \label{fig:argmultcc}
\end{figure} 

\begin{figure}
   \centering
   \includegraphics[width=0.95\columnwidth]{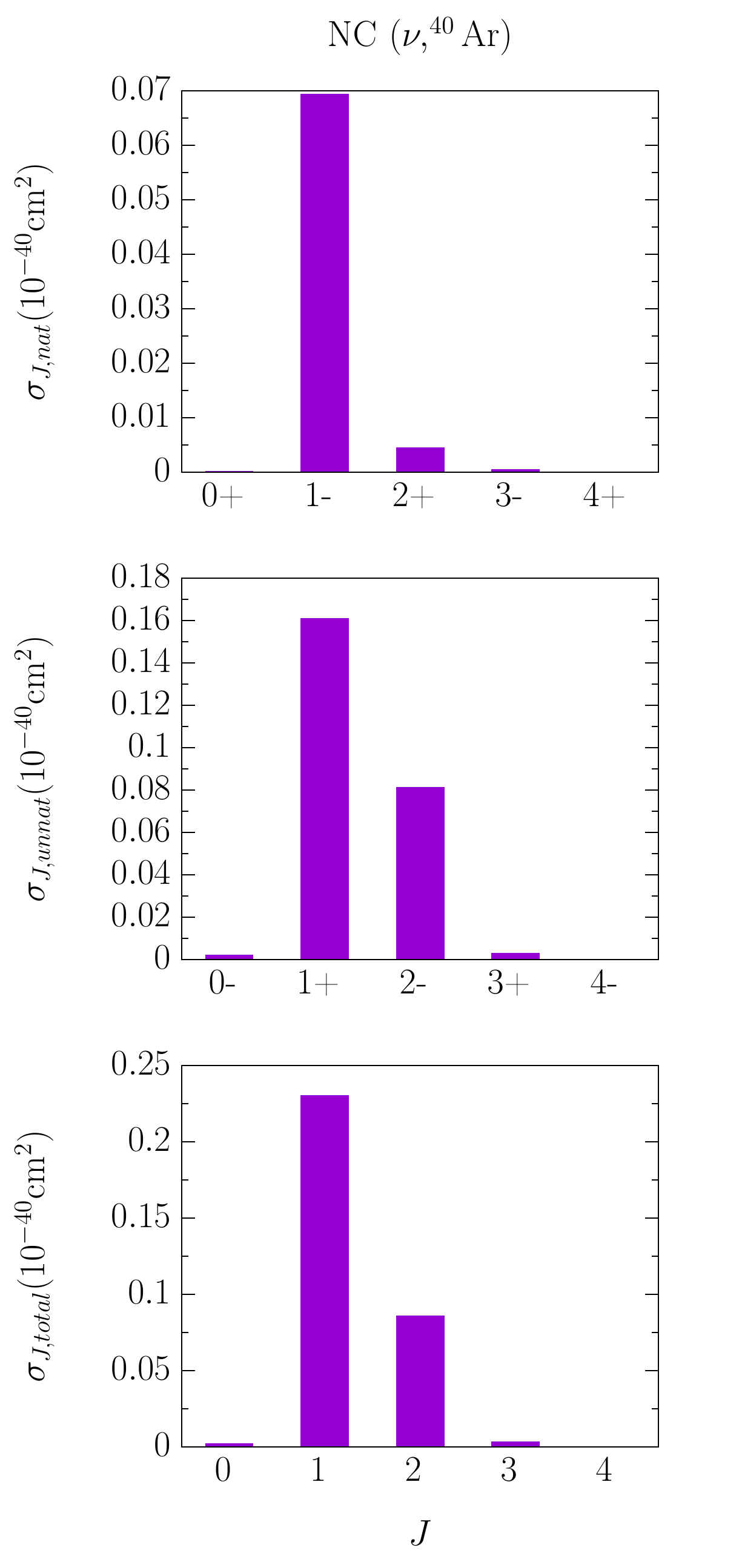}
   \caption{Contributions of different multipole transitions to neutral--current reactions for a DAR $\nu_e$ spectrum.}
   \label{fig:argmultnc}
\end{figure} 

\begin{figure}[!hb]
   \centering
   \includegraphics[width=0.95\columnwidth]{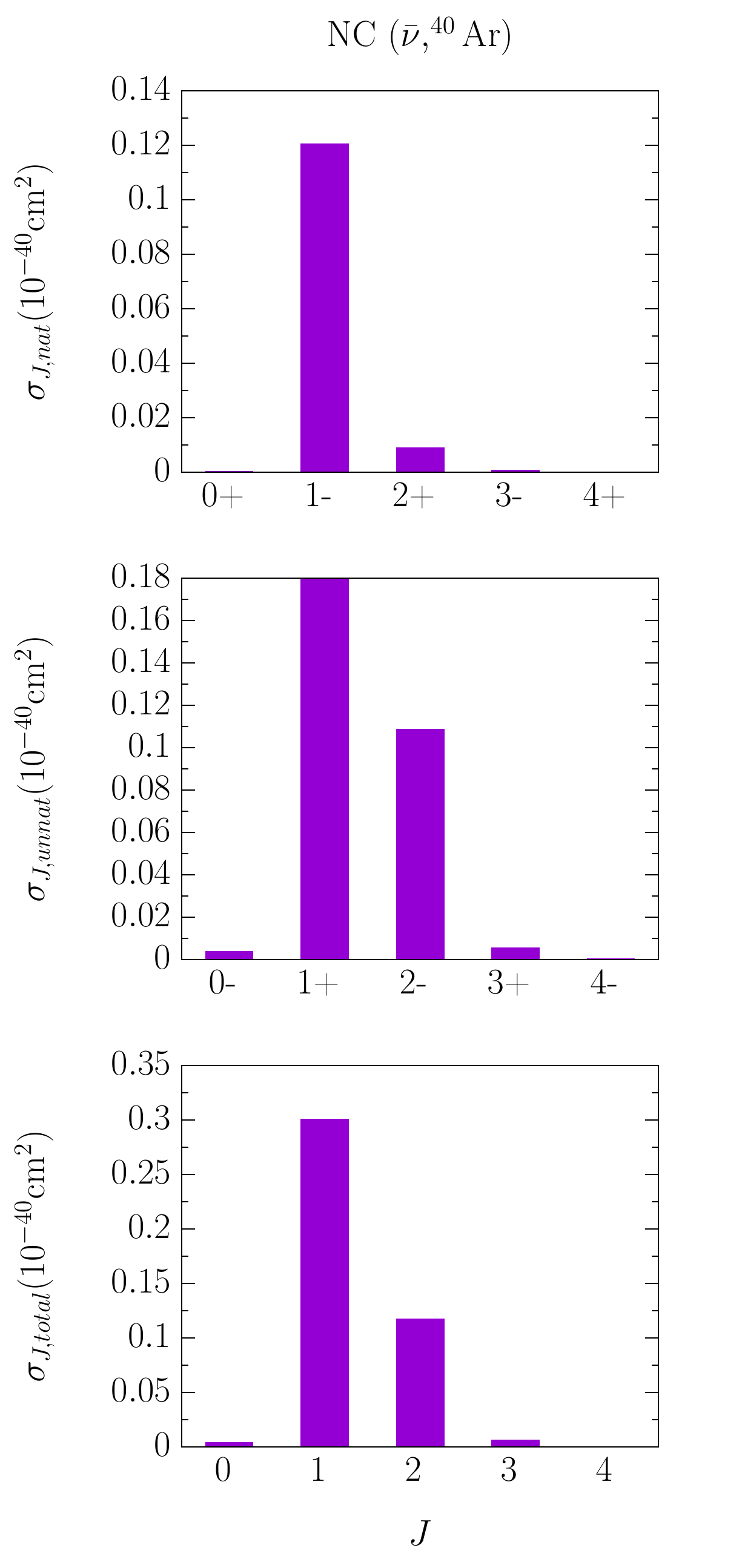}
   \caption{Contributions of different multipole transitions to neutral--current reactions for a DAR $\bar{\nu}_\mu$ spectrum.}
   \label{fig:argmultncanti}
\end{figure} 

We now provide HF+CRPA model predictions for neutrino--${}^{40}$Ar cross sections at low incoming neutrino energies. We first present total cross sections as a function of incoming energy. In figures \ref{fig:argtotcc}, \ref{fig:argtotnc}, \ref{fig:argtotccanti} and \ref{fig:argtotncanti} we show the reaction strength for CC and NC, for both incoming $\nu_e$ and $\bar{\nu}_e$.

In the plots the total cross sections are shown for different multipole cutoffs $J$. We also show the contribution of natural ($0^+,1^-,2^+,3^-$, ...) and unnatural ($0^-,1^+,2^-,3^+$, ...) parity transitions separately. In comparing the case of CC and NC for neutrinos and antineutrinos, other than the overall magnitudes of the cross sections, the qualitative features concerning the relative contributions from different multipolarities are similar. The $0^+$ and $1^+$ transitions correspond with the allowed approximation. One can appreciate the non--trivial contributions arising from the forbidden $1^-$ and $2^-$ transitions in the continuum channels. The importance of these two in particular was similarly noted in Ref.~\cite{Kostensalo:2018kgh}. Even higher order multipoles start becoming visible above 50 MeV. The $0^+$ reaction strength is negligible for excitations into the continuum. In comparing the CC and NC result, one notices that the difference in strength between neutrinos and antineutrinos is noticeably larger for the former. A key reason for this is the Coulomb interaction with the outgoing lepton, which enhances or decreases the cross section for neutrinos and antineutrinos, respectively. \\

Next, we cover the energy dependence of the contributions of different multipole moments. We first take a look at the differential cross section $\frac{\mathrm{d}\sigma}{\mathrm{d}\omega}$ for ${}^{40}$Ar in Fig.~\ref{fig:singlediffarg} for several incoming energies. For clarity, we do not show the $0^-$ and $0^+$ contributions, since these are small. At 25 MeV incoming energy, we can see that the cross section is quite well described in an allowed approximation, with the differential cross section almost coinciding with the $1^+$ transition strength, except for a small difference in the tail. At 50 MeV, this is no longer an adequate approximation, with significant strength coming from the $1^-$ and $2^-$ transitions, affecting the shape of the excitation spectrum as well. Finally, at 75 MeV, even higher order multipolarities become non-negligible. In Figs.~\ref{fig:ctotcc} and \ref{fig:fetotcc} we investigate the A--dependence of the cross sections. Comparing Figs.~\ref{fig:argtotcc} to \ref{fig:ctotcc} and \ref{fig:fetotcc}, one observes that even for a relatively small nucleus as ${}^{12}$C, forbidden transitions are non--negligible at higher energies, but not to nearly the same degree as they are for ${}^{40}$Ar and ${}^{56}$Fe, both of which show a qualitatively similar picture, not differing too significantly in size. In general, the larger the target nucleus, the more multipoles will be necessary for convergence of the cross section.

In order to appreciate the effect that the individual multipole contributions will have in an experimental context, it is instructive to take a look at folded cross sections. For now, we make use of the spectra displayed in Fig.~\ref{fig:michel}, yielded by $\pi^+$ pions decaying at rest (DAR), with associated fluxes
\begin{equation}\label{eq:fluxnue}
n(E_{\nu_e}) = \frac{96E_{\nu_e}^2}{m_\mu^4}(m_\mu - 2E_{\nu_e}),
\end{equation}
and
\begin{equation}\label{eq:fluxnumu}
n(E_{\bar{\nu}_\mu}) = \frac{32E_{\bar{\nu}_\mu}^2}{m_\mu^4}\left(\frac{3}{2}m_\mu - 2E_{\bar{\nu}_\mu}\right),
\end{equation}
with $ E_\nu \in \left[ 0 , m_\mu/2 \right]$ and $m_\mu \approx$ 105.7 MeV the rest mass of a muon. We fold the cross sections on display in figures \ref{fig:argtotcc}, \ref{fig:argtotnc}, \ref{fig:argtotccanti} and \ref{fig:argtotncanti} (except for CC ($\bar{\nu}_\mu,{}^{40}$Ar), which is not kinematically available) with the appropriate fluxes of Eqs.~\ref{eq:fluxnue} and \ref{eq:fluxnumu}. The results, plotted individually for each multipole, are shown in Figs.~\ref{fig:argmultcc}, \ref{fig:argmultnc} and \ref{fig:argmultncanti}. The plots demonstrate that for argon nuclei, forbidden transitions can be expected to contribute sizeably to the reaction strength in the energy regimes of DAR neutrinos. All three paint a qualitatively similar picture in terms of relative contributions, except for the case of NC ($\bar{\nu}_\mu,{}^{40}$Ar), where the $\bar{\nu}_\mu$ flux peaks at a higher value, as seen in Fig.~\ref{fig:michel}. Therefore the $2^-$ transition strength has a larger relative weight. And although small, even the $2^+$, $3^-$ and $3^+$ transitions cannot be neglected when high precision is desired. We conclude that working strictly in an allowed approximation is not desirable, and forbidden transitions must be properly accounted for in the modeling of low--energetic neutrino--nucleus scattering events. Therefore it is worth noting that the Monte Carlo generator MARLEY~\cite{marley} employs Fermi and Gamow--Teller transitions in its modeling of charged--current neutrino--${}^{40}$Ar scattering, but does not go beyond this~\cite{Gardiner2017}. In light of our results, the implementation of forbidden transitions should be considered important.

\section{Summary}\label{sec:summ}

Argon represents an important nuclear scattering target in low--energy neutrino research. We have presented calculations for ($\nu,{}^{40}$Ar) scattering. We employed a Hartree Fock + Continuum Random Phase Approximations (HF+CRPA) framework, which allows us to model the responses and include the effects of long--range correlations. The Coulomb interaction of the outgoing lepton is accounted for in an effective way by interpolating between the Fermi function, valid at low outgoing charged lepton momenta, and the Modified Effective Momentum Approach (MEMA) at higher ones. Subsequently comparing model results for ${}^{56}$Fe nuclei, one notices that the model predictions compare favorably. Finally, we presented our predictions for ${}^{40}$Ar, for both neutral and charged current events, where the important role played by forbidden transitions is evident. Even in a typical experimental scenario of a $\pi^+$ decay--at--rest neutrino spectrum, this is still the case, with strong contributions from the $1^-$ and $2^-$ multipoles, and smaller contributions for higher multipolarities.

 \begin{acknowledgments}
   This work was supported by the Research Foundation Flanders (FWO--Flanders). The computational resources (Stevin Supercomputer Infrastructure) and services used in this work were provided by the VSC (Flemish Supercomputer Center), funded by Ghent University, FWO and the Flemish Government – department EWI.
 \end{acknowledgments}

 \bibliography{biblio}
 \end{document}